# Discretized Volumes in Numerical Methods

Miklós Antal* and Mihály Makai**

* BME antalmi@gmail.com
**BME Institute of Nuclear Techniques, Department of Nuclear Techniques, Budapest, Hungary, makai@reak.bme.hu

**Abstract**

We present two techniques novel in numerical methods. The first technique compiles the domain of the numerical methods as a discretized volume. Congruent elements are glued together to compile the domain over which the solution of a boundary value problem is sought. We associate a group and a graph to that volume. When the group is symmetry of the boundary value problem under investigation, one can specify the structure of the solution, and find out if there are equispectral volumes of a given type. The second technique uses a complex mapping to transplant the solution from volume $V_1$ to volume $V_2$ and a correction function. Equation for the correction function is given. A simple example demonstrates the feasibility of the suggested method.

**1. Problem description**

In both science and engineering, we solve boundary values in a volume composed of large number of meshes. This is the case in nuclear engineering[1], in fluid dynamics[2,3] and electromagnetic fields[4]. We address the question: under what conditions are the solutions for two meshes identical, or, are the solutions transformable into each other by a simple rule? How to find the transformation rule? How can we find equivalent meshes?

In the design and safety analysis of large industrial devices, calculational models are tested against experiments carried out on a small scale mock-up. This is the case with nuclear power plants[5], aeroplanes[6a], ships[6b]. We would need a transplantation of the measured values to the geometry of the real scale device. Is there any hope of doing that exactly or we have to put up with approximate methods[7]?

We suggest two methods for attacking the above mentioned problems.

**Method 1. Discretized volumes**. The domain $V$, over which the solution of the boundary value problem is sought is constructed by the following procedure. We choose an appropriate tile $t$, and glue copies of the tile by their corresponding sides. Although in principle $t$ is almost arbitrary[8], we confine the discourse to triangular shaped tiles. Since triangulation is a well known and widely used technique even in theoretical problems[9], this is not considered as a limitation. The discretized volumes considered by us are always finite. A concise description of the discretized volume $V$ is a finitely presented group $G$ [10]. Although in practical problems the applications of computational group theory are rather limited, the authors strongly hope for a steady development in both computational tools (software) and means (hardware). So, a discretized volume is described by the tile $t$ and the group $G$. A further asset, a graph $\mathcal{G}$ is also defined. If copies $i$ and $j$ of tile $t$ are interconnected by an edge $\alpha$ of $t$, graph vertices $i$ and $j$ are also interconnected by an edge $s_\alpha$. Analyzing the group $G$, the graph $\mathcal{G}$, one can easily reveal basic properties of $V$.

**Method 2. Complex mapping**. It is well known that every simply connected planar domain can be mapped to the unit disk by a complex function[11]. Unfortunately most of those maps do not commute with the operator in the differential equation to be solved, hence, if we seek a solution on the image of $V$, we have to introduce a correction function to satisfy the same equation we had in $V$. We derive an equation for the correction function. When the BVP has a unique solution (up to normalization), the correction function can be determined thus yielding a transplantation rule.

The main results of the present work are:

1. We provide an algebraic description of the discretized volume. By analyzing the group and graph associated with a given discretized volume, one can answer a number of questions.
2. Using that algebraic description, we can formulate conditions for isospectral volumes to exist.
3. We can formulate a formal solution.
4. We give conditions for two discretized volumes to be isospectral and we show that the eigenfunctions of isospectral discretized volumes are transformed into each other by a linear map.
5. We provide a transplantation recipe based on complex mapping.

In order to demonstrate the feasibility of item 5, we present an example: the neutron diffusion equation is solved in a disk and we transplant the result into a square.

The structure of the manuscript is as follows. We define the discretized volume in Section 2, along with the associated group theoretic assets, group and graph. In Section 3, we formulate the formal solution of the boundary value problem and fix the conditions on the existence of isospectral discretized volumes. Section 4 is devoted to the complex map and the correction function. In Section 5, we deal with a possible application: code scaling. Concluding remarks are given in Section 6.

**2. Algebraic description of discretized volumes (DV)**

In the discourse, we are dealing with two-dimensional (2D) problems. The reason is the following. Let us consider the map $V_1 \rightarrow V_2$. By cutting $V_1$ and $V_2$ with a plane, perpendicular to the $z$ axis, we get two, two-dimensional domains $D_1$ and $D_2$ marked by blue and red colors. If we find a map $f : D_1 \rightarrow D_2$ shown in Fig. 1, we can construct the map $f(z) : V_1 \rightarrow V_2$.

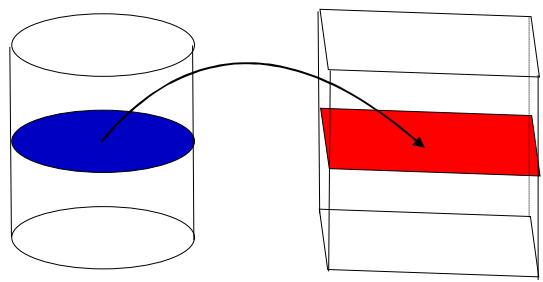

*Fig. 1. Mapping a three-dimensional region into another*

## 2.1. Discretized Volumes (DV)

**Definition 2.1.** A tile $t$ is a finite, simply connected domain of the plane $\mathbb{R}^2$ bordered by three finite straight lines called sides. Every two sides have a common point called corner.

**Definition 2.2.** A discretized volume $V$ is composed of $N < \infty$ (finite number) copies of tile $t$ so that we stick copies of tile $t$ to each other along a corresponding side. Those copies of $t$, which share a side are called adjacent. The shared side is called internal. When $N > 1$, every copy of $t$ has at least one adjacent side.

**Definition 2.3.** The discretized volumes $V_1, V_2, \ldots$ are arranged into equivalent classes. $V_1$ and $V_2$ are in the same class if there is a finite sequence of rotations, translation, and reflections that maps $V_1$ into $V_2$, or vice versa. Equivalent discretized volumes are denoted as $V_1 \sim V_2$.

**Definition 2.4.** Let us denote the sides of $t$ by $\alpha$, $\beta$, and $\gamma$. A graph $\mathcal{G}$ is assigned to $V$ in the following way. We number the copies of $t$ in $V$. If copies numbered $n_1$ and $n_2$ are adjacent, and they share a side of type $\alpha$, then the vertices $n_1$ and $n_2$ of graph $\mathcal{G}$ are connected by an edge of type $\alpha$.

**Definition 2.5.** We associate a permutation group $G$ to $V$ in the following way. When in $V$ a side of type $\alpha$ connects the copies $i_{\alpha 1}$ with $i_{\alpha 2}$ and $j_{\alpha 1}$ with $j_{\alpha 2}$ ... then, we form the permutation $a = (i_{\alpha 1}, i_{\alpha 2})(j_{\alpha 1}, j_{\alpha 2})\ldots$. We carry out that procedure for sides $\alpha$, $\beta$ and $\gamma$ to get generators $a$, $b$, and $c$, and group $G$ is generated by $a$, $b$, and $c$.

**Definition 2.6.** The action ($g \circ t$) of $g \in G$ on a tile $t$ is defined as follows. Since $G$ is a finitely presented group, every $g \in G$ is a product of the generators. Hence, we have to define the action of the generators on $t$. Let $\alpha$ be a side of $t$, then $a \circ t$ is a copy of $t$ obtained by reflecting $t$ through side $\alpha$. If the length of $g$ is $n$ then $g \circ t$ is a discretized volume of $(n+1)$ copies of $t$.

**Definition 2.7.** The action of $g \in G$ on $t$ is given by Definition 2.6. The orbit of $t$ under the group $G$ is the set $g \circ t$ for all $g \in G$.

**Remark.** Definitions 2.6 and 2.7 are needed solely in the construction of DV.

**Definition 2.8.** The action $g \circ V$ of $g \in G$ on $V$, consisting of $N$ copies of $t$, is defined as follows. Let $a$ be a generator of $G$. The action of $a$ on copy $i$ is $a \circ i = i$ whenever side $\alpha$ of copy $i$ is not internal in $V$. Otherwise $a \circ i = j$ if copies $j$ and $i$ ($i, j < N$) share an internal side of type $\alpha$.

**Remark.** With a given tile $t$, $N$ may be limited.

**Definition 2.9.** Adjacency matrix $\mathbf{A}_V$ of $V$ is an $N \times N$ matrix, its $(i,j)$ element is 1 if copies $i$ and $j$ are adjacent, otherwise it is zero.

**Definition 2.10.** The auxiliary matrix **X** is **D+A**, where **D** is a diagonal matrix, its $i^{th}$ entry equals the number of internal sides of copy $i$ in $V$.

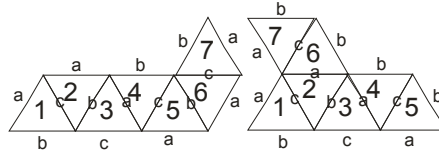

*Fig 2. Two discretized volumes composed of 7 copies of a regular triangle*
*(a,b, and c should read as $\alpha$, $\beta$ and $\gamma$)*

Two simple discretized volumes are shown in Fig. 2. Each volume is comprised of seven copies of a regular triangle tile.

**Lemma 2.11.** If $V_1 \sim V_2$ and the corresponding graphs are $\mathcal{G}_1$ and $\mathcal{G}_2$, then $\mathcal{G}_1 \sim \mathcal{G}_2$.
*Proof:* There is an isomorphism between the copies and edges of the two discretized volumes, this entails the statement. □

**Lemma 2.12.** Let the number of copies of $t$ in discretized volume $V$ be $N$. Then, in group $G$ associated with $V$, there is a subgroup of index $N$. Consequently, the order of group $G$ is a multiple of the number of copies ($N$) in $V$.
*Proof:* There is a map between finite groups and Cayley graphs. A coset representation of G is isomorphic with graph $\mathcal{G}$, in which there are N vertices. □

In accordance with Lemma 2.11 $\mathcal{G}_1 \sim \mathcal{G}_2$ suffices for $V_1 \sim V_2$. In accordance with Lemma 2.12, $G_1 \sim G_2$ suffices for $V_1 \sim V_2$.

*2.2. Equispectral discretized volumes*

Let us investigate the following eigenvalue problem in $V_1$:

$$\mathbf{A}\Phi(x) = \lambda\Phi(x), \ x \in V_1 \tag{1}$$

where operator **A** is such that

$$[\mathbf{A},\mathbf{O}] = \mathbf{AO} - \mathbf{OA} = 0. \tag{2}$$

Here **O** is an automorphism of $V_1$, viz. a reflection, translation or rotation operator acting on functions defined over $V_1$. We assume that reflections are symmetries of **A**. Eq. (2) is the usual definition of symmetries of operator **A**. The set of operations with which $V_1$ is formed involves only symmetries of operator **A**. Definitions 2.6 and 2.8 assure group $G$ to be isomorphic to a group of transformations commuting with **A**. It is well known that the solution of eigenvalue problem (1) with a homogeneous boundary condition along the boundary $\partial V_1$ of $V_1$ is easily determined from a solution of the same problem over another member of the class $\sim V_1$ (see Definition 2.3). The question is, if we can find a volume $V_2$ not equivalent to $V_1$ such that all the eigenvalues of problem (1) will remain the same as for $V_1$.

**Definition 2.13.** If there exist volumes $V_1$ and $V_2$, which are not equivalent according to definition 2.3 such that all the eigenvalues of problem (1) are the same in $V_1$ and $V_2$, we call $V_1$ and $V_2$ equispectral.

The benefit from knowing equispectral volumes comes from the fact that it is rather tiresome to solve Eq. (1) even for a simple volume. At the same time we know that equivalent volumes provide an easily feasible recipe for transplanting the solution from one member of the class to another. Knowledge of equispectral volumes would widen the range of transformations where instead of solving Eq. (1) over a new equispectral volume, one would apply a relatively simple transformation to an already known solution.

Actually, in a number of practical problems, one would be satisfied with the equivalence of one eigenvalue, in other words, with the equality of the eigenvalue and a transformation rule for the eigenfunctions. In a number of cases physical meaning is attributed to the so called fundamental mode eigenvalue.

### 3. Solution of a Boundary Value Problem over DVs

*3.1. The formal solution*

Below we derive a formal solution to problem (1) with Dirichlet boundary condition (i.e. Φ=0 on the boundary). The solution is given in terms of the Green's function $G_t$ of tile $t$, that we obtain as the solution of the following boundary value problem:

$$(\mathbf{A} - \lambda)G_t(x,\xi) = 0, \quad x \in t \tag{3}$$

$$G_t(x,\xi) = \delta(x-\xi), \quad \xi \in \partial t. \tag{4}$$

Solution of

$$\mathbf{A}\Phi(x) = \lambda \Phi(x), \quad x \in t \tag{5}$$

with boundary condition

$$\Phi(x) = \begin{cases} f_a(\xi), \xi \in \partial t_a, 0 \text{ for } \xi \in \partial t_b, \partial t_c \\ f_b(\xi), \xi \in \partial t_b, 0 \text{ for } \xi \in \partial t_a, \partial t_c \\ f_c(\xi), \xi \in \partial t_c, 0 \text{ for } \xi \in \partial t_b, \partial t_a \end{cases}, \tag{6}$$

where the three sides of tile $t$ are $\partial t_a$, $\partial t_b$, and $\partial t_c$ is

$$\Phi(x) = \int_{\partial t_a} G_t(x,\xi)f_a(\xi)d\xi + \int_{\partial t_b} G_t(x,\xi)f_b(\xi)d\xi + \int_{\partial t_c} G_t(x,\xi)f_c(\xi)d\xi, \quad x \in t. \tag{7}$$

Since $V$ consists of copies of $t$, the solution of Eq. (1) in $V$ is sum of integrals like Eq. (7). Let us assume that there are $K$ internal sides (cf. Definition 2.2) and the solution is $f_i(\xi)$ along side $i$. We compose Φ(x) from $N$ functions $\underline{F}(x) = (\Phi_1(x),...,\Phi_N(x))$, one from each copy of $t$. Then,

$$\underline{F}(x) = \mathbf{Q} \cdot \underline{v}(x), \tag{8}$$

where

$$v_k(x) = \int_{\partial t_k} G_t(x,\xi) f_k(\xi) d\xi, \quad k = 1,\ldots,K. \tag{9}$$

Matrix $\mathbf{Q}$ is $N \times K$ [1], and assigns sides and copies of $t$. As we see, expression (8) has two components, $\underline{v}(x)$ depends only on tile $t$ and operator $\mathbf{A}$, hence we call it the physical part of the solution. On the other hand, $\mathbf{Q}$ depends only on the structure of $V$, hence we call it the structural part of the solution. Structure (8) can be used to derive an integral equation set for the solution along the internal boundaries and serves as basis for approximate solution methods. Now we are interested in the structural part.

**Lemma 3.1.** The auxiliary matrix $\mathbf{X}$ of volume $V$ and the structural matrix $\mathbf{Q}$ in Eq. (8) are related as $\mathbf{X}=\mathbf{QQ}^+$.

*Proof:* $X_{ij}$ is the scalar product of lines $i$ and $j$. Hence, $X_{ii}$ equals the number of internal sides of copy $i$. Element $X_{ij}$ is 1 if copy $i$ and $j$ share an internal side, and zero otherwise. This is just the definition of $\mathbf{X}$. □

**Lemma 3.2.** Let $\mathbf{X}_i$ be the auxiliary matrix of discretized volume $V_i$, $i=1,2$. If the eigenvalues of matrices $\mathbf{X}_i$ are the same then $V_1$ and $V_2$ are isospecral.

*Proof:* There exist matrix $\mathbf{M}$ such that $\mathbf{X}_2=\mathbf{MX}_1\mathbf{M}^+$, and $\mathbf{X}_i=\mathbf{Q}_i\mathbf{Q}_i^+$. Let $\mathbf{O}$ be a $K \times K$ orthogonal matrix, then $\mathbf{X}_2=\mathbf{Q}_2\mathbf{Q}_2^+= \mathbf{MX}_1\mathbf{M}^+=\mathbf{MQ}_1\mathbf{OO}^+\mathbf{Q}_1^+ \mathbf{M}^+$. This entails $\mathbf{Q}_2=\mathbf{MQ}_1\mathbf{O}$, using this in Eq. (8) for $V_2$, we get: $F_2(x)=\mathbf{MQ}_1\mathbf{O}\,v_2(x)$. Let $\mathbf{O}$ be such that $\mathbf{O}v_2=v_1$, and we get $F_2(x) =\mathbf{MQ}_1\,v_1(x) =\mathbf{M}\,F_1(x)$. This expression transforms each and every eigenfunction on $V_1$ into the corresponding eigenfunction on $V_2$, independently from $\lambda$, hence $V_1$ and $V_2$ are isospectral. □

Lemma 3.2. asserts that algebraic tools ascertaining similarity of the auxiliary matrices lead to find equispectral volumes. Furthermore, features of equispectral volumes guaranteed by Lemma 3.2 can be fixed as follows.

**Lemma 3.3.** Let $V_1$ and $V_2$ be equispectral volumes according to Lemma 3.2. Then

1. $V_1$ and $V_2$ are composed of the same tile $t$.
2. In $V_1$ and $V_2$ the number of copies ($N$) of $t$ are equal.
3. Along the external boundary of $V_1$ and $V_2$ the number of sides are equal by side tipes.
4. Along the internal boundaries of $V_1$ and $V_2$ the number of sides are equal by side tipes.

*Proof:* The transplantation rule does not affect the component of the solution therefore the tile used in $V_1$ and $V_2$ is the same. Since matrix $\mathbf{M}$ is a square matrix, item 2 is trivial. The transplantation rule leads to transformed functions on the internal boundaries but the transformation corresponds to the orthogonal matrix $\mathbf{O}$, which leaves the number, and type of internal boundary unchanged, therefore items 3 and 4 follow.

---

[1] Since every reflection creates an internal side, K=N-1.

With a triangular tile $t$, the copies of $t$ in discretized volume $V$ fall into two categories (say black and white) as we can color $V$ by two colors. Let $\underline{w}$ be an $N$-tuple with elements +1 or -1 in position $i$ when copy $i$ is black or white, respectively.

**Lemma 3.4.** The auxiliary matrix of the above considered discretized volume $V$, has the following property: $\mathbf{X}\underline{w}=0$. The solution $\underline{F}(x)$ of problem (1) has the following property: either $\underline{F}(x) = \Phi(x)\underline{w}$ where $\Phi(x)$ is the solution of problem (1) on the tile $t$, or $\underline{F}(x)\underline{w}=0$.
*Proof:* The first part of the statement is an immediate consequence of the structure of the auxiliary matrix. The second part of the statement is a particular case of Hersch's theorem[12].

*3.2. Constructing equispectral volumes*

The present Section is devoted to the problem of finding equispectral volumes. Our analysis is based on $G$ and $\mathcal{G}$, the group and graph associated with discretized volume $V$. Let us start with the investigations initiated by Sunada[13-20]. Let $g \in G$ and let $\{g\}$ denote the conjugacy class of $g$ in $G$. Let $G_1, G_2 \subset G$ subgroups in $G$. We say $(G, G_1, G_2)$ to form a Sunada triple if the number of elements from subgroups $G_1$ and $G_2$ are the same in every $\{g\}$. Let $M$ be a manifold.

**Sunada theorem.** Let $V$ be a compact Riemannian manifold, $G$ a finite group acting on $M$ by isometries. Suppose that $(G, G_1, G_2)$ is a Sunada triple, and that $G_1$ and $G_2$ act freely on $M$. Then the quotient manifolds $M_1 = G_1 \backslash M$ and $M_2 = G_2 \backslash M$ are isospectral.

Starting from Sunada theorem, Gordon and Webb[16] showed how to construct planar regions in which the Laplace operator is isospectral.
One can find equispectral discretized volumes $V_i$, $i=1,2$ by the Sunada theorem so that one searches for Sunada triples. To this end, we have to find in the group $G$ associated with $V_1$, $V_2$ two subgroups of index $N$. The GAP program[21] offers means to solve that task. In Appendix A, we show such an algorithm[2]. The result is two coset representations of $G$ in terms of subgroups $G_1$ and $G_2$, from that we construct $V_1$ and $V_2$.

Following the recipe by Gordon and Webb[16], we study the graphs associated with $V_1$ and $V_2$. In that, the following conjecture is utilized.

**Conjecture 3.5.** Let $V_1$ and $V_2$ be equispectral volumes with associated graphs $\mathcal{G}_1$ and $\mathcal{G}_2$ and associated groups $G_1$ and $G_2$. If graphs $\mathcal{G}_1$ with edges[3] $s_i$ and $\mathcal{G}_2$ with edges $v_i$ are isomorphic, and the isomorphism of the edges is $s_i = g_i v_i g_i^{-1}$, then $V_1 \sim V_2$ does not hold unless all $g_i$-s are automorphisms of tile $t$. $G_1$ and $G_2$ are isomorphic groups.

Buser[19] has presented isospectral planar graphs which are not isomorphic but the associated discretized volumes are not planar ones. Buser's graphs are derived from abstract groups and the adjacency proposed by the coset representation exclude the associated geometry to be planar. In Appendix A, we give a proof of Conjecture 3.5 for a particular case.

---

Since isomorphic graphs contain the same number of vertices of degrees 1, 2, and 3, one seeks isospectral discretized volumes by specific features (length of walks, number of vertices of degree 3) of the graph. To this end, we have analyzed $V$ composed of seven regular triangles. There are 25 non-isomorphic graphs, the order of the associated groups are 5040=7!, 2520, and 168. The classification is given in Appendix B. Equispectral volumes exist within a family of non-isomorphic graphs provided the tile $t$ is not symmetric. We show an example first presented in Ref. [13]. The order of either associated group is 2520, the generators of the DV on the left are $a = (4,6)(5,7)$, $b = (3,5)(2,4)$ and $c = (1,2)(5,6)$, whereas the generators of the DV on the right are $a' = (3,7)(2,6)$, $b' = (3,5)(2,4)$, $c' = (1,2)(5,6)$. The generators are related as $c' = c$, $b' = b$ and $a' = aba$. The associated graphs are $1 - c - 2 - b - 4 - a - 6 - c - 5 <_{b-3}^{a-7}$ (left) and $7 - a' - 3 - b' - 5 - c' - 6 - a' - 2 <_{c'-1}^{b'-4}$ (on the right). (Here $i - c - j$ stands for vertices $(i, j)$ connected by an edge of type $c$. A vertex of degree 3 has two connecting walks.) The algorithm worked out by Zhen-yun Peng and Ya-xin Peng in Ref. [22] comes rather handy in the search of isospectral volumes. They studied the general problem

$$\mathbf{AXB} + \mathbf{CYD} = \mathbf{E} \qquad (10)$$

where $\mathbf{A}$, $\mathbf{B}$, $\mathbf{C}$, and $\mathbf{E}$ are given, one of $\mathbf{X}$ and $\mathbf{Y}$ is the sought rectangular matrix. This analysis confirms the results in Ref. [16], and the $\mathbf{M}$ and $\mathbf{O}$ matrices, associated with the problem in the previous paragraph, with the notation of Lemma 3.2 are

$$\mathbf{M} = \begin{pmatrix} 0 & 1 & 1 & 0 & 1 & 0 & 0 \\ 1 & -1 & 0 & 0 & 0 & 0 & 0 \\ 1 & 0 & -1 & 0 & 0 & 0 & 1 \\ 0 & 1 & 0 & 0 & 0 & -1 & -1 \\ 1 & 0 & 0 & 0 & -1 & 1 & 0 \\ 0 & 0 & 0 & -1 & 1 & 0 & 1 \\ 0 & 0 & 1 & -1 & 0 & -1 & 0 \end{pmatrix}$$

$$\mathbf{O} = \begin{pmatrix} 0 & 0 & 0 & 1 & 0 & -1 \\ 0 & 1 & 0 & 0 & 1 & 0 \\ 1 & 0 & 1 & 0 & 0 & 0 \\ 0 & 0 & 0 & 1 & 0 & 1 \\ -1 & 0 & 1 & 0 & 0 & 0 \\ 0 & 1 & 0 & 0 & -1 & 0 \end{pmatrix}$$

In accordance with Conjecture 3.5, not equivalent equispectral volumes would not contain perfectly symmetric tiles, hence equispectral graphs should be sought within a given category, with some less regular tile form.

In the analysis of discretized volumes of practical calculations, the presented method may serve only as a theoretical investigation for two reasons. The first one is the large number of copies in a discretized volume. The algorithm presented in Appendix C works only for $N < 10$, which is really far from the practical applications ($N > 1000$). The second one is in the limited choice of equispectral volumes: they have the same number of tiles, the external and internal boundary types should correspond in the two DVs. Some of the limitations may be relaxed or removed in the future.

## 4. Mappings

The transplantation given in Section 3 is fairly general. When the involved operator is linear, one can find another transplantation technique. This technique, however, applies only to a given one dimensional eigenspace, i.e. when the λ eigenvalue is associated with a one dimensional eigenspace. We assume that the λ eigenvalue is known. Details are elucidated below.

**Theorem 4.1.** Let us consider the BVP in domain $V_1$ as

$$\mathbf{A}\Psi(\underline{x}) = \lambda \Psi(\underline{x}), \ \underline{x} \in V_1 \tag{4.1}$$

and boundary condition

$$\Psi(\underline{x}) = 0, \ \underline{x} \in \partial V_1, \tag{4.2}$$

and in domain $V_2$ as

$$\mathbf{B}\Phi(\underline{x}') = \lambda \Phi(\underline{x}'), \ \underline{x}' \in V_2 \tag{4.3}$$

and boundary condition

$$\Phi(\underline{x}') = 0, \ \underline{x}' \in \partial V_2. \tag{4.4}$$

Here **A** and **B** are linear operators, and **B** is obtained by the $x \rightarrow x'$ substitution in operator **A**. Let the one-to-one map **T**: $V_1 \rightarrow V_2$ be given. Then

$$\Phi(\underline{x}') = [\mathbf{T}(\Psi)](\underline{x}') + g(\underline{x}') \tag{4.5}$$

and the correction function $g(\underline{x}')$ meets the following equation:

$$(\mathbf{B} - \lambda)g(\underline{x}') = -Q(\underline{x}'), \ \underline{x}' \in V_2, \tag{4.6}$$

and the source $Q(\underline{x}')$ is given by

$$Q(\underline{x}') = (\mathbf{B} - \lambda)[\mathbf{T}(\Psi)](\underline{x}'). \tag{4.7}$$

*Proof:* Substituting (4.5) into Eq. (4.3) and rearranging it, one obtains (4.6) and (4.7). □

In the example given below, we show how to transform the solution of the one energy group neutron diffusion equation ($\Delta f - k^2 f = 0$), where $k$ is a constant, over a square region ($V_2$, coordinates: $\underline{x}' = (x', y')$, solution is known to be $\Psi(x')$) into the solution on a disk ($V_1$, coordinates $\underline{x}=(x,y)$, solution is sought and denoted by $\Phi(\underline{x})$). Since both solutions are analytical, the accuracy of the method can readily be tested.

The transformation relating the unit disk and a square region with unit sides can be found with the Scwarz-Christoffel method[11]. We start from the disk (shown in Fig. 3) and apply the following transformation to get the square (apparently, variables $x$ and $y$ are treated as the real and immaginary part of a complex variable $z$):

$$T[z] = 1. + (0. + 0.539353\,\mathrm{i})\sqrt{(1-\mathrm{i}) + \frac{2}{-\mathrm{i}+z}}\sqrt{(1+\mathrm{i}) - \frac{2\,\mathrm{i}}{1+z}}\;\mathrm{EllipticF}\!\left[\mathrm{i}\,\mathrm{ArcSinh}\!\left[\frac{1}{\sqrt{-\frac{(1+\mathrm{i})\,(\mathrm{i}+z)}{-1+z}}}\right],\,2\right]$$

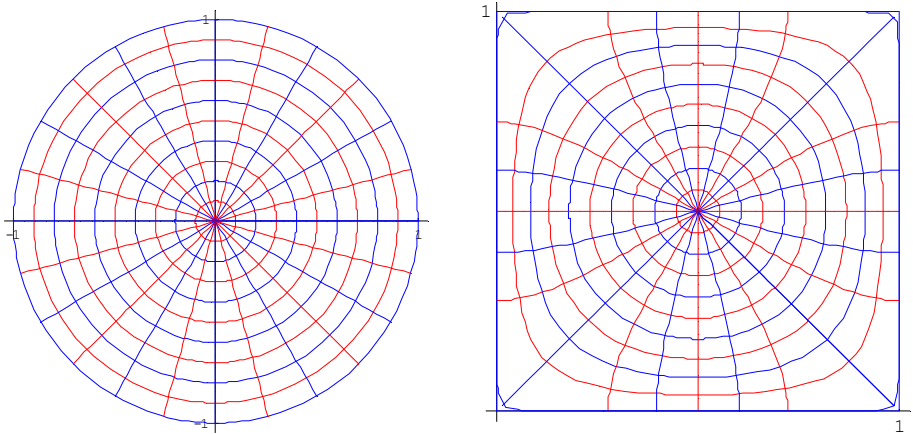

Fig 3. The unit disk and the unit square. Colors indicate transformed surfaces.

We know the solution on the square to be composed of normalized sinus functions:

$$\Psi(\underline{x}') = \sqrt{\frac{2}{l}}\cdot\sin\!\left(\frac{\pi\cdot\mathrm{Re}[x'+iy']}{l}\right)\cdot\sqrt{\frac{2}{m}}\cdot\sin\!\left(\frac{\pi\cdot\mathrm{Im}[x'+iy']}{m}\right)$$

Thus:

$$[\mathbf{T}(\Psi)](\underline{x}) = \sqrt{\frac{2}{l}}\cdot\sin\!\left(\frac{\pi\cdot\mathrm{Re}[\mathbf{T}(x'+iy')]}{l}\right)\cdot\sqrt{\frac{2}{m}}\cdot\sin\!\left(\frac{\pi\cdot\mathrm{Im}[\mathbf{T}(x'+iy')]}{m}\right)$$

With polar coordinates ($x' = r'\cdot\cos(\varphi')$ and $y' = r'\cdot\sin(\varphi')$):

$$[\mathbf{T}(\Psi)](\underline{x}) = \sqrt{\frac{2}{l}}\cdot\sin\!\left(\frac{\pi\cdot\mathrm{Re}[\mathbf{T}(r'\cos\varphi'+ir'\sin\varphi')]}{l}\right)\cdot\sqrt{\frac{2}{m}}\cdot\sin\!\left(\frac{\pi\cdot\mathrm{Im}[\mathbf{T}(r'\cos\varphi'+ir'\sin\varphi')]}{m}\right)$$

The original solution on the square and its transformed form is depicted in Fig. 4.

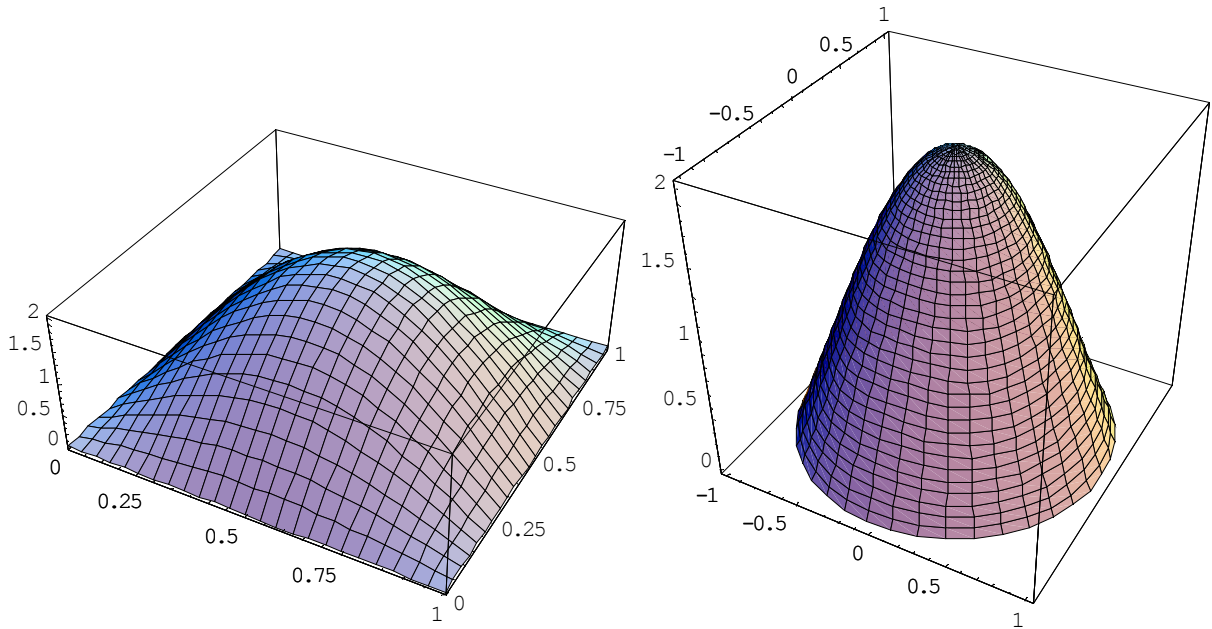

*Fig. 4. Solution on the unit square and its transformed version on the disk.*

It is easy to calculate the correction term because an analytical solution can be given on the disk as well: any solution is composed of first order Bessel functions. Expanding **T(Ψ)** with the Bessel functions (Bessel zeros are the multipliers in the brackets) we get:

$$[\mathbf{T}(\Psi)](\underline{x}) = -2.0062 J_0(2.40483r) + 0.00894 J_0(5.52008r) - 0.00381 J_4(7.58834r) -$$
$$- 0.00054 J_4(11.0674r) - 0.00049 J_8(12.2251r) - 0.00013 J_8(16.0378)$$

This approximation is fairly precise: the difference between the analytical solution on the disk and the function transplanted from the disk ($\mathbf{T}(\Psi)$) is 0,14% in case of expansion with 6 terms. Though, this difference is definitely not a numerical error: $\mathbf{T}(\Psi)$ is only composed of functions with four fold rotational symmetry that is not characteristic to the disk. This difference is contained in the correction term $g(x)$.

**5. Code scaling made different**

In studying large industrial devices mock-ups, scaled down devices have been used for a long time. The measurements are transferred to the real size facility by a scaling law. Such scaling recipes are more than a century old. In hydrodynamics, the basic equations are non-linear [24], therefore the scaling is restricted to specific „characteristic" distances, velocities etc. Various scaling methods are used even today, for example in the nuclear industry[25] and in code validation[26].

In subsection 5.1 we formulate the code scaling or transplantation problem and in subsection 5.2 outline solutions based on the results of the previous sections.

*5.1. The problem*

The principal problem of code scaling is formulated as follows. We need a model for a large and expensive device. To study the relevant physical processes, to verify the performance of the model we carry out measurements on a scaled down model. In a basic report of the nuclear industry[27], the motivation is formulated in the following way: *"It is obvious that full scale experiments will be too dangerous and costly, thus it is necessary to use scale models and simulation experiments."* Similarity principles are utilized in the design of the experimental scaled down facility and also in the transplantation of the measurements to the real devise. As Ishii and Kataoka write[27] : *"Similarity laws and scaling criteria are quite important for designing, performing, and analyzing simulation experiments using a scale model."*

The mathematical aspect of the problem is formulated as follows. The physical processes are governed by an equation:

$$\mathbf{A}(p)\Psi(x,p) = \lambda \Psi(x,p), \ x \in V_1 \qquad (5.1)$$

where operator $\mathbf{A}(p)$ depends on a set of parameters $p$, acts on the space variables $x$, and the physical process is characterized by $\Psi(x,p)$. We assume that Eq. (5.1) unequivocally determines the state function $\Psi(x,p)$. The device is finite, at its boundary a suitable, homogeneous boundary condition is prescribed.

We formulate the problem of code scaling as follows. We are looking for a domain $V_2$ such that

$$\mathbf{B}(x')\Phi(x',q) = \lambda \Phi(x',q), \ x' \in V_2. \qquad (5.2)$$

We assume the existence of a one-to-one map $\mathbf{T}: V_1 \rightarrow V_2$. Operator $\mathbf{B}(q)$ acts on space variable $x'$ and is obtained from $\mathbf{A}(p)$ by introducing $x = f(x')$ with the help of map $\mathbf{T}$. The parameters are also affected by the map $\mathbf{T}$ because diameters, lengths (geometrical data in general) are also among the parameters. We need a transplantation rule $\Psi(x,q) \rightarrow \Phi(x',p)$ to transplant the measured values from $V_1$ to $V_2$.

*5.2. The transplantation*

The results presented in sections 2-4 propose the following solutions of the transplantation problem.

*5.2.1. Transplantation by DVs*

The discretized volumes (DVs) introduced in Section 2, suggest the following solution of the transplantation problem. If we subdivide volumes $V_1$ and $V_2$ into congruent domains, for example triangles, used in Section 2, so that both of them are DVs, we can apply the technique presented in Section 3.

**Conjecture 5.1.** Let $V_1$ and $V_2$ be DVs constructed from the same tile $t$, let either volume be composed from $N$ copies of $t$. Let $V_1$ and $V_2$ be inequivalent DVs, with associated groups $G_1$ and $G_2$ and associated graphs $\mathcal{G}_1$ and $\mathcal{G}_2$. Then, by analyzing the associated groups $G_1$ and $G_2$, as well as the associated graphs $\mathcal{G}_1$ and $\mathcal{G}_2$, it is possible to ascertain if $V_1$ and $V_2$ are equispectral or not.

If $G_1 \sim G_2$, there is a chance for $G_1$ and $G_2$ to be isospectral. To verify that, we have to express the generators $a_2, b_2, c_2$ of $G_2$ by the generators $a_1, b_1$ and $c_1$ of $G_1$. In accordance with Lemma 2.12, there must be two subgroups of index $N$ in $G_2$ (and in the isomorphic $G_1$ as well). Let these subgroups be $H_1$ and $H_2$. If $(G, H_1, H_2)$ is a Sunada triple, $V_1$ and $V_2$ are isospectral by the Sunada theorem. It is not known if there are other cases when $V_1$ and $V_2$ are isospectral.

**Theorem 5.2.** Let $\mathbf{X}_1$ and $\mathbf{X}_2$ be the auxiliary matrices of DVs $V_1$ and $V_2$. Let $\Psi(x, p) = (\psi_1(\xi), \psi_2(\xi), \ldots, \psi_N(\xi))$ and $\Phi(x', q) = (\phi_1(\xi), \phi_2(\xi), \ldots, \phi_N(\xi))$ be the solutions of (5.1) and (5.2), respectively so that the $i^{th}$ element of the respective vector is the solution in the $i^{th}$ tile of $V_1$ and $V_2$, respectively. $\xi$ is the coordinate in tile $t$. Then, the formula

$$\phi_i(\xi) = \sum_{j=1}^{N} S_{ij} \psi_j(\xi), \ \xi \in t \quad (5.3)$$

transplants the solution from $V_1$ to $V_2$ provided $\mathbf{SX}_1\mathbf{S}^+ = \mathbf{X}_2$.

*Proof:* Theorem 5.2 is immediate consequence of Lemma 3.2.

*5.2.2. Transplantation by conformal mapping*

In a number of physical problems we are interested in a specific eigenvalue problem, to which belongs a one dimensional eigenspace. Then, the complex maps are applicable in solving the transplantation problem.

**Theorem 5.3.** Let $V_1$ and $V_2$ be given, let the eigenvalue $\lambda$ of problem (5.1) be known. Let $\mathbf{T}: V_1 \to V_2$ be given. Let the measured values of the eigenfunction $\Phi$ be known on $V_1$. Then, the transplanted measured values in $V_2$ are given by

$$\Phi = \mathbf{T}\Psi + f \quad (5.3)$$

where

$$(\mathbf{B} - \lambda)f = Q \quad (5.4)$$

and

$$Q = -(\mathbf{B} - \lambda)\mathbf{T}\Psi. \quad (5.5)$$

*Proof:* We have to repeat the steps in the proof of Theorem 4.1. A formal solution is

$$f = \sum_{i \neq k} c_i \varphi_i \quad (5.6)$$

where

$$\mathbf{B}\varphi_i = b_i \varphi_i \quad (5.7)$$

and $b_k = \lambda$.

# 6. Concluding remarks

Let us start with a list of open problems.
1. It has not yet been proven that, derived from Sunada's theorem, the solutions of problem (1) with appropriate boundary conditions on equispectral volumes obey Lemma 3.2.
2. It is a question if Cayley graphs of planar equispectral volumes are always isomorphic (Conjecture 3.5).
3. Whether there is an algorithm to solve the problem considerably faster than with the algorithm in Appendix A?
4. Allows that speed for solving practical problems (N>1000)?
5. Buser, Conway, Doyle, and Semmler expressed their guess: equispectral volumes are rather scarce. Is it true for very large DVs (i.e. for $N \approx 10^5$)?
6. Can the formal solution (8) be generalized for DVs with different tiles?

We achieved the following results:

1. Formulated a general boundary value problem for DVs and associated an algebraic description with the DVs, viz. a graph and a group.
2. By analysing the group and the graph, one can establish relations between boundary value problems on specific DVs.
3. We improved the formal solution given in Ref. [23]. The improvements have lead to Lemma 3.4 and Conjecture 3.5.
4. Conjecture 3.5 leads to a classification of possible DVs. We made the classification of the DVs composed of seven triangles, the most frequently encountered problem in the literature [13-19], [23]. Known equispectral DVs obey our observations.
5. We presented a method for transplanting a one dimensional eigenspace of the eigenvalue problem using complex mapping and a correction function.
6. We presented an application of the above method.

The code scaling originates from simple transformations of the units in which basic physical quantities (length, time, and mass) are measured. In thermal hydraulics, where the measurements are carried out on a *model* and are used to assess the accuracy of the calculation of a *real device*, the method can be summarized as follows[28]:

1. The basic conservation equations (material, momentum and energy conservations) are put down. We also need the state equation which involves the pressure, material density, and enthalpy. The equations are put down for the *model* and the *real device*.
2. The *model* and the *real device* are characterized by the following parameters:

    a. ratio of characteristic lengths
    b. ratios of characteristic time, velocity and acceleration
    c. ratio of heat generation per unit volume.

3. By comparing the conservation equations for the model and the *real device*, a set of relationships is obtained among the parameters mentioned in item 2.

As a conclusion, at the end of their work, Nahavandi, Castellana and Moradkhanian[29] arrive at three scaling laws called time reducing, and two versions of time preserving scales. The scaling law is a set of expressions[27] to be preserved by the model (for details which are of interest only for experts in heat and mass transfer see Ref.[27]), among others:

- Richardson number
- Friction number
- Modified Stanton number
- Biot number
- Reynolds number.

When the *model*, which is a scaled down experimental facility, and the *real device*, which might be as complex as a nuclear power plant, have identical Richardson number, friction number etc. one may hope for a transplantation recipe. But this is not the case. The recipe provided by code scaling is global in the sense that transplants properly only overall features of the thermal hydraulical phenomena, not the detailed solution as misbelieved by some. Although the derived simple relations have proved rather useful, for example in designing static states of *model*s[27] but the transplantation of detailed solution has not been achieved. The difficulty is in the non-homogeneous nature of the phenomena under investigation. In a part of the device water, in another steam, in a third one a mixture of steam and water flows. Those phenomena can not be transplanted by a single rule.

The invariance properties of the involved equations have been studied[30,31] for almost two decades but only the advent of computer codes has made it possible to extend the set of problems being studied. Such analyses have revealed that the symmetry groups of the conservation equations, at least in their Boussinesq form[32], have only the Euclidean group E(2) as their symmetry in plane geometry. No trace of the groups derived by the scaling method has been found. No wonder, since code scaling formulates not precise invariance observed at each point but general relationships.

But there are other criticisms as well conserning code scaling. Madrid and Alhama[27] formulates the following objection to the classical dimension analysis „*…for the problems of interest in engineering, it provides more dimensionless monomials (with no clear physical significance) than strictly necessary for the solution.*"

The methods we presented in Section 5, represent only the first steps towards a new kind of code scaling.

Discretized volumes (DVs) show a way in which non-equivalent geometries can be found to solve an equation and a simple transplantation rule is also given. Then the transplantation is exact. It is true that in the present form the procedure is simple and of restricted use. However, there are reserves to be exploited. Triangles that can be transformed into each other by a linar map may allow for an extension of the method.

As to complex maps, it is a long computation to solve the equation for the correction function. Only numerical procedures are available to realize mapping between two complex regions. It is a difficult job to create a complex map bringing a given domain to another given domain. More complex forms may be studied by using the Rvachov technique[33]. The results presented here, are meant as the first demonstrative steps to point out: there are new possibilities in the application of algebraic methods in code scaling.

# Appendix A. A proof of Conjecture 3.5 for a particular case

Although the general proof of Conjecture 3.5 is not known, and there are opinions that the statement may not hold in general, it is necessary to show that for our purposes the conjecture can be used.

**Statement A.1.** A Cayley-graph is given ($\mathcal{G}_A$), which is depicted with isosceles triangle nodes ($N$), the associated discretized volume (DV) is denoted by $A$. When the equal sides of a node are $a$ and $c$, and the third side is denoted by $b$, then a change of $a$ and $c$ in the graph means a reflection of the discretized volume: $\mathbf{T}A = B$. (The graph we get after the change is $\mathcal{G}_B$. The reflection is denoted by $T$, and $B$ is the discretized volume we get after the change in the graph.)

*Proof:* For the proof we use complete induction. If we start DV $\mathcal{G}_A$ and $\mathcal{G}_B$ at the same node ($N_0$) fixed at a position, then the statement holds for this zeroth order case: the node is symmetric to the altitude perpendicular to the base: $\mathbf{T}M_0 = M_0$. Side $a$ and $c$ are mirror images of each other, side $b$ is reflected to itself.
Let us denote the sides of triangle $N_i$ by $a_i, b_i, c_i$. Thus: $\mathbf{T}a_0^A = c_0^B, \mathbf{T}b_0^A = b_0^B, \mathbf{T}c_0^A = a_0^B$.

Then, in each step a new triangle ($N_i$) is obtained and drawn in both DVs by reflecting a formerly got triangle. We show, that if the statement is true for step $n$ (that means $\mathbf{T}A_n = B_n$), then it holds for step ($n+1$) (meaning $\mathbf{T}A_{n+1} = B_{n+1}$).
We distinguish two cases:
If reflection $t$ (here $t$ denotes a reflection to a side and it affects only one triangle, when $\mathbf{T}$ is the reflection to the ordinary axis of symmetry and it affects the whole DV) is made with respect to side $b$, then we are reflecting to such sides in the two DVs, that are mirror images of each other: $\mathbf{T}b_i^A = b_i^B$. It also holds for the actually reflected triangles: $\mathbf{T}N_i^A = N_i^B$ (as $\mathbf{T}A_i = B_i \;\; \forall i \leq n$-re). Using Lemma 2.11, the statement follows. It is subsequently true, that: $\mathbf{T}a_i^A = c_i^B, \mathbf{T}b_i^A = b_i^B, \mathbf{T}c_i^A = a_i^B \;\; \forall i$.
If the reflection is made with respect to side $a$ or $c$, then besides $\mathbf{T}N_i^A = N_i^B$ we can also observe that $\mathbf{T}a_i = c_i$ and $\mathbf{T}c_i = a_i$, so Lemma 2.11 can be used as well. So: $\mathbf{T}a_i^A = c_i^B, \mathbf{T}b_i^A = b_i^B, \mathbf{T}c_i^A = a_i^B \;\; \forall i$. □

**Lemma A2.** Let us consider a reflection with respect to a certain ($a$, $b$ or $c$) side (action of operator **t**). If $\mathbf{T}N_i^A = N_i^B$ and $\mathbf{T}\mathbf{t}^A = \mathbf{t}^B$ then $\mathbf{T}(\mathbf{t}N_i^A) = \mathbf{t}N_i^B$, where $\mathbf{t}^A$ and $\mathbf{t}^B$ denote the reflection axis in DV $A$ and $B$ respectively.
*Proof:* the reflection of a triangle can be described with the reflection of its vertices.
Two vertices of $N_i^A$ are on the reflection axis, so they are transformed by **T** into the appropriate vertices in the other graphic. Third vertices are mirror images of each other, as it is easily it can be seen after basic coordinate geometrical considerations. (Reflection $t$ should be considered as a coordinate geometrical transformation, then it is trivial, that **T** transforms those vertices into each other.) So: $\mathbf{T}a_i^A = c_i^B, \mathbf{T}b_i^A = b_i^B, \mathbf{T}c_i^A = a_i^B$. □

**Appendix B. Classifications of DVs composed of seven regular triangles**

| No. | Discretised Volume | N3* | NISB** Order of G | Graph | Generators |
|---|---|---|---|---|---|
| 1. | 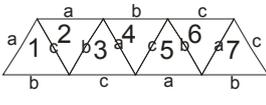 | 0 | 2,2,2 2520 | 1-c-2-b-3-a-4-c-5-b-6-a-7 | a: (3,4),(6,7) b: (2,3),(5,6) c: (1,2),(4,5) |
| 2. | 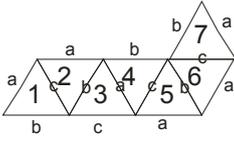 | 0 | 1,2,3 5040 | 1-c-2-b-3-a-4-c-5-b-6-c-7 | a: (3,4) b: (2,3),(5,6) c: (1,2),(4,5),(6,7) |
| 3. | 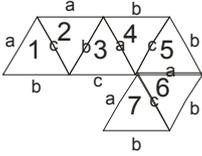 | 0 | 2,1,3 5040 | 1-c-2-b-3-a-4-c-5-a-6-c-7 | a: (3,4),(5,6) b: (2,3) c: (1,2),(4,5),(6,7) |
| 4. | 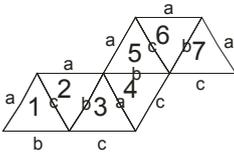 | 0 | 1,3,2 5040 | 1-c-2-b-3-a-4-b-5-c-6-b-7 | a: (3,4) b: (2,3),(4,5),(6,7) c: (1,2),(5,6) |
| 5. | 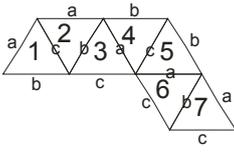 | 0 | 2,2,2 2520 | 1-c-2-b-3-a-4-c-5-a-6-b-7 | a: (3,4),(5,6) b: (2,3),(6,7) c: (1,2),(4,5) |
| 6. | 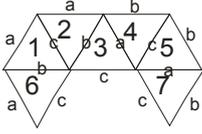 | 0 | 2,2,2 2520 | 6-b-1-c-2-b-3-a-4-c-5-a-7 | a: (3,4),(5,7) b: (2,3),(1,6) c: (1,2),(4,5) |
| 7. | 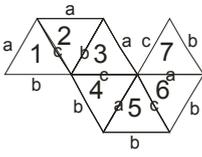 | 0 | 2,1,3 | 1-c-2-b-3-c-4-a-5-c-6-a-7 | a: (4,5),(6,7) |
|     |     |     | 5040 |     | b: (2,3) |
|     |     |     |      |     | c: (1,2),(3,4)(5,6) |
| 8.  | 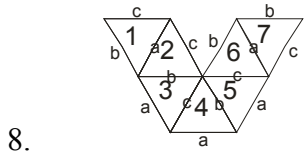 | 0 | 2,2,2 | 1-a-2-b-3-c-4-b-5-c-6-a-7 | a: (1,2),(6,7) |
|     |     |     | 2520 |     | b: (2,3),(4,5) |
|     |     |     |      |     | c: (3,4),(5,6) |
| 9.  | 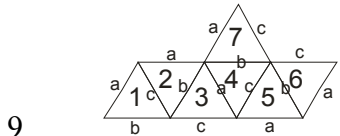 | 1 | 1,3,2 | $1-c-2-b-3-a-4<^{b-7}_{c-5-b-6}$ | a: (3,4) |
|     |     |     | 5040 |     | b: (2,3),(4,7),(5,6) |
|     |     |     |      |     | c: (1,2),(4,5) |
| 10. | 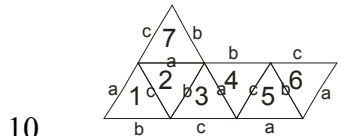 | 1 | 2,2,2 | $1-c-2<^{a-7}_{b-3-a-4-c-5-b-6}$ | a: (2,7),(3,4) |
|     |     |     | 168  |     | b: (2,3),(5,6) |
|     |     |     |      |     | c: (1,2),(4,5) |
| 11. | 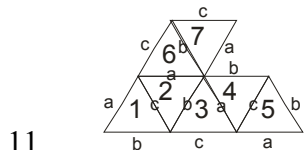 | 1 | 2,2,2 | $1-c-2<^{a-6-b-7}_{b-3-a-4-c-5}$ | a: (2,6),(3,4) |
|     |     |     | 2520 |     | b: (2,3),(6,7) |
|     |     |     |      |     | c: (1,2),(4,5) |
| 12. | 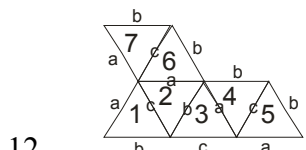 | 1 | 2,1,3 | $1-c-2<^{a-6-c-7}_{b-3-a-4-c-5}$ | a: (2,6),(3,4) |
|     |     |     | 5040 |     | b: (2,3) |
|     |     |     |      |     | c: (1,2),(4,5),(6,7) |
| 13. | 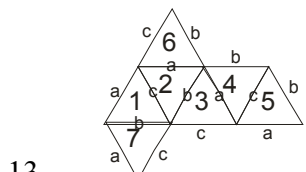 | 1 | 2,2,2 | $7-b-1-c-2<^{a-6}_{b-3-a-4-c-5}$ | a: (2,6),(3,4) |
|     |     |     | 168  |     | b: (2,3),(1,7) |
|     |     |     |      |     | c: (1,2),(4,5) |
| 14. | 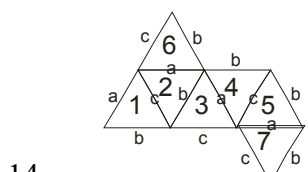 | 1 | 3,1,2 | $1-c-2<^{a-6}_{b-3-a-4-c-5-a-7}$ | a: (2,6),(3,4),(5,7) |
|     |     |     |     | 5040 | b: (2,3) |
|     |     |     |     |      | c: (1,2),(4,5) |

15. 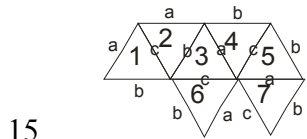   1   2,1,3   $1-c-2-b-3<^{a-4-c-5-a-7}_{c-6}$   a: (3,4),(5,7)

       5040                                                                          b: (2,3)

                                                                                         c: (1,2),(3,6),(4,5)

16. 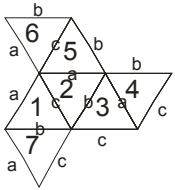   1   2,2,2   $7-b-1-c-2<^{a-5-c-6}_{b-3-a-4}$   a: (2,5),(3,4)

       168                                                                            b: (1,7),(2,3)

                                                                                          c: (1,2),(5,6)

17. 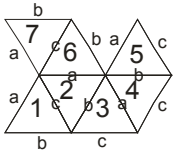   1   2,2,2   $1-c-2<^{a-6-c-7}_{b-3-a-4-b-5}$   a: (2,6),(3,4)

       2520                                                                        b: (2,3),(4,5)

                                                                                          c: (1,2),(6,7)

18. 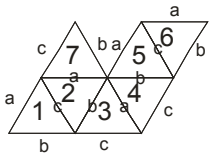   1   2,2,2   $1-c-2<^{a-7}_{b-3-a-4-b-5-c-6}$   a: (2,7),(3,4)

       2520                                                                       b: (2,3),(4,5)

                                                                                          c: (1,2),(5,6)

19. 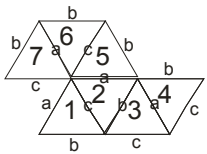   1   3,1,2   $1-c-2<^{a-5-c-6-a-7}_{b-3-a-4}$   a: (2,5),(3,4),(6,7)

       5040                                                                       b: (2,3)

                                                                                          c: (1,2),(5,6)

20. 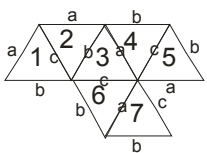   1   2,1,3   $1-c-2-b-3<^{c-6-a-7}_{a-4-c-5}$   a: (3,4),(6,7)

       5040                                                                       b: (2,3)

                                                                                          c: (1,2),(3,6),(4,5)

| | | | | | |
|---|---|---|---|---|---|
| 21. 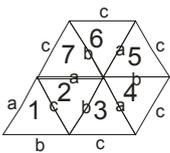 | | 1 | 3,3,1<br>5040 | $1-c-2<^{a-7-b-6-a-5}_{b-3-a-4-b\lrcorner}$ | a: (2,7),(3,4),(5,6)<br>b: (2,3),(4,5),(6,7)<br>c: (1,2) |
| 22. 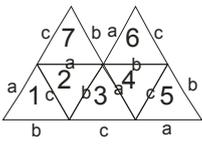 | | 2 | 2,2,2<br>2520 | $1-c-2<^{a-7}_{b-3-a-4}\ \angle^{b-6}_{c-5}$ | a: (2,7),(3,4)<br>b: (2,3),(4,6)<br>c: (1,2),(4,5) |
| 23. 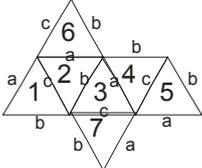 | | 2 | 2,1,3<br>5040 | $1-c-2<^{a-6}_{b-3}\ \angle^{a-4-c-5}_{c-7}$ | a: (2,6),(3,4)<br>b: (2,3)<br>c: (1,2),(3,7),(4,5) |
| 24. 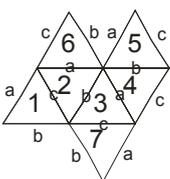 | | 2 | 2,2,2<br>2520 | $1-c-2<^{a-6}_{b-3}\ \angle^{a-4-b-5}_{c-7}$ | a: (2,6),(3,4)<br>b: (2,3),(4,5)<br>c: (1,2),(3,7) |
| 25. 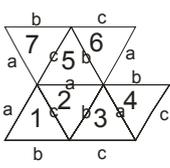 | | 2 | 2,2,2<br>2520 | $1-c-2<^{b-3-a-4}_{a-5-----}\ \angle^{c-7}_{b-6}$ | a: (2,5),(3,4)<br>b: (2,3),(5,6)<br>c: (1,2),(5,7) |

a, b and c should read as $\alpha$, $\beta$ and $\gamma$
N3*: Number of nodes of degree 3
NISB**: Number of Internal Sides by Boundary Type and order of the associated group G

## Appendix C. GAP algorithm to find Equispectral volumes

The GAP program given below has been written to find discretized volumes $V_i$ isospectral to a given discretized volume $V_0$. $V_0$ is defined by its three generators. The discretized volumes $V_i$ are characterized by the number of copies connected by sides of type $a$, $b$ and $c$. The version presented here has been made to verify if the program finds the discretized volumes of Gordon and Webb[16].

```
#
#Discretized volume V is given by the three generators of group G.
# Now V is made up from 7 copies ot tile t
#
# The search starts from V0 given also by three generators d, e and f
#
 d:=(3,7)(2,6); e:=(2,4)(3,5); f:=(1,2)(5,6);
#
LogTo("Tri7");
G:=Group(d,e,f);
Size(G);
CT:=[];; ri:=[];
cc:=ConjugacyClassesSubgroups(G);;
ccr:=List(cc,x->Representative(x));;
ccf:=Filtered(ccr,x->Size(x)*7=Size(G));;
#
# In ccf, we have collected all subgroups of index 7
# Next we apply generators d, e, f, to them
#
for i in [1..Size(ccf)] do
Append(CT,CosetTableBySubgroup(G,ccf[i]));Append(ri,[i]);
od;
#
# In CT, we have the discretized volumes represented by coset tables
# (One line is associated with ech generator and its inverse)
#
```

**Acknowledgement**

The kind assistance of Dr. Erzsébet Lukács in writing the GAP algorithm is acknowledged.